\documentstyle[pre,aps,epsf,amssymb]{revtex}

\newcommand{\be}{\begin{equation}}
\newcommand{\ee}{\end{equation}}
\newcommand{\bi}{\begin{itemize}}
\newcommand{\ei}{\end{itemize}}

\newcommand{\disp}{\displaystyle}

\newcommand{\vxi}{{\mathbf \xi}}
\newcommand{\Zeta}{{\mathbf x}_{\zeta}}
\newcommand{\x}{{\mathbf x}}

\newcommand{\y}{{\mathbf y}}
\newcommand{\vl}{{\mathbf l}}

\newcommand{\J}{{\mathbf J}}
\newcommand{\R}{{\mathbf R}}
\newcommand{\C}{{\mathbf C}}
\newcommand{\W}{{W}}
\newcommand{\A}{{\widehat{A}}}
\newcommand{\tA}{{\widetilde{A}}}
\newcommand{\tG}{{\widetilde{G}}}
\newcommand{\B}{{\widehat{B}}}
\newcommand{\tB}{{\widetilde{B}}}

\newcommand{\oprho}{{\widehat{\rho}}}
\newcommand{\oprhot}{{\widehat{\rho_t}}}
\newcommand{\trho}{{\widetilde{\rho}}}
\newcommand{\mH}{{\mathbf H}}

\newcommand{\opL}{\widehat{L}}

\newcommand{\oq}{\widehat{q}}
\newcommand{\op}{\widehat{p}}

\newcommand{\mA}{{\mathbf A}}
\newcommand{\mP}{{\mathbf P}}
\newcommand{\mD}{{\mathbf D}}
\newcommand{\M}{{\mathbf M}}

\newcommand{\zero}{{\mathbf 0}}

\newcommand{\tW}{\widetilde{W}}

\newcommand{\tr}{\mathop{Tr}}

\newcommand{\ch}{\mathop{ch}}

\newcommand{\der}{\partial}

\renewcommand{\Re}{{\rm Re}}

\begin{document}

\title{Symplectic evolution of Wigner functions in Markovian open systems}

\author{O. Brodier\footnote{brodier@cbpf.br}, 
A. M. Ozorio de Almeida\footnote{ozorio@cbpf.br}}

\address{Centro Brasileiro de Pesquisas Fisicas, 
Rua Xavier Sigaud 150, 22290-180, 
Rio de Janeiro, R.J., Brazil.}

\maketitle

\begin{abstract}

The Wigner function is known to evolve classically under the exclusive action
of a quadratic Hamiltonian. If the system also interacts with the environment
through Lindblad operators that are complex linear functions of position and 
momentum, then the general evolution is the convolution of a 
nonhamiltonian classical
propagation of the  Wigner function with a phase space Gaussian that 
broadens in time.
We analyze the consequences of this in the three generic cases of elliptic, 
hyperbolic and parabolic
Hamiltonians. The Wigner function always becomes positive in a definite time, 
which does not depend on the initial pure state. We observe the influence of 
classical dynamics and dissipation upon this threshold.  
We also derive an exact formula 
for the evolving linear entropy as the average of a narrowing Gaussian
taken over a probability distribution that depends only on the initial
state. This leads to a long time asymptotic formula for the growth of
linear entropy. We finally discuss the possibility of recovering the initial
state.

\end{abstract}

\section{Introduction}

The correspondence between classical and quantum mechanics of closed 
dynamical systems is most perfect for quadratic Hamiltonians. In this case, 
the classical evolution is linear, like its quantum counterpart, and 
generates an orbit within the group of symplectic (linear, canonical) 
transformations in phase space \cite{Arn:book}. These are directly 
linked to the corresponding quantum metaplectic group \cite{Vor77}. Indeed, 
the evolution operator in any of the usual representations is merely the 
complex exponential of the classical generating function \cite{Lit86}. Of 
course, quadratic Hamiltonians are a very special case, but they include 
the ubiquitous harmonic oscillator, the parabolic potential barrier and 
the free particle, which form adequate starting points for the analysis 
of more complex motion.

 The Weyl representation of an arbitrary quantum operator $\A$ is
\be
A(\x) \equiv  \int dq' \langle q+\frac{q'}{2} | \A 
| q-\frac{q'}{2} \rangle \exp{\left(-i\frac{pq'}{\hbar}\right)},
\label{wignerdef}
\ee
that is, $\A$ is represented in phase space, $\x = (p,q) $, by the Weyl 
symbol $A(\x)$. The Wigner function $W(\x)$ is then the Weyl symbol for 
$\oprho/2\pi\hbar$, where $\oprho$ is the density operator. Like all Weyl 
symbols, the Wigner function propagates classically under the action of 
a quadratic Hamiltonian \cite{Vor77}\cite{Lit86}:
\be
\frac{\der}{\der t} W_t(\x) = \left\{ H(\x), W_t(\x) \right\},
\ee
introducing the classical Poisson bracket on the right side \cite{Arn:book}
and $H(\x)=\x\cdot\mH\x$, the Weyl symbol of the quadratic Hamiltonian. 
The symbol $\cdot$ stands for the inner scalar product.
Hence one has $W_t(\x)=W_0(\R_{-t}\x)$, where 
\be
\R_t=\exp{\left(2\J\mH t\right)}
\label{defRt}
\ee
is the $2\times 2$ matrix giving the classical Hamiltonian time evolution
of a phase space point $\x$.
Actually, this propagation is also shared by the Fourier transform
of $W_t(\x)$,
\be
\tA(\vxi) = \frac{1}{2\pi\hbar} \int d\x ~ 
\exp{\left(-\frac{i}{\hbar}\vxi\wedge\x\right)} A(\x),
\label{chordtrans}
\ee
i.e. it is also true that
\be
\frac{\der}{\der t} \tW_t(\vxi) = \left\{ H(\vxi), \tW_t(\vxi) \right\}.
\ee
Note that $H(\vxi)$ must be read literally as the classical Hamiltonian
$H(\x)$ taken at the point $\vxi=(\xi_p,\xi_q)$, which is in general different
from the chord transform $\widetilde{H}(\vxi)$ of $H(\x)$.
In other words, one has also $\tW_t(\vxi) = \tW_0 (\R_{-t}\vxi)$.
Above we have made use of the wedge product,
\be
\vxi\wedge\x \equiv \xi_p q - \xi_q p \equiv (\J\vxi)^T\x \equiv 
\J\vxi \cdot \x,
\label{wedgedef}
\ee
also defining the transpose of a vector $(.)^T$ and the matrix $\J$.
The semiclassical 
background for the symplectic invariance of both the Wigner function 
and its Fourier transform is that the Weyl phase space coordinate $\x$ may be 
associated to pairs of points in phase space, $\x_{\pm}$, by 
$\x=(\x_++\x_-)/2$.
The conjugate variable to this centre is the chord $\vxi = \x_+-\x_-$. 
The linear motion of both the chord $\vxi$ and the center $\x$ is the same 
as for each individual phase space point 
$\x_+$ or $\x_-$. We will refer to $\tW(\vxi)$ as the chord 
function as in \cite{Alm98}, though it is also known as the characteristic 
function in quantum optics.

The question that we address is to what extent can the simplicity and 
generality of symplectic motion of closed quantum systems be incorporated 
within the description of systems whose coupling to the environment cannot be 
ignored. In this case the evolution is no longer unitary, unless the full 
Hamiltonian of the system combined with the environment is taken into account. 
All the same, a certain measure of generality is restored by the assumption 
that the density operator is governed by a Markovian master equation 
\cite{Lin76},
\be
\frac{\der\widehat{\rho}}{\der t}=-\frac{i}{\hbar}
\bigl[\widehat{H}, \widehat{\rho}\bigr]
-{1\over{2\hbar}}\sum_j ~
2 \widehat{L}_j \widehat{\rho} \widehat{L}^{\dagger}_j 
- \widehat{L}^{\dagger}_j \widehat{L}_j \widehat{\rho}
- \widehat{\rho} \widehat{L}^{\dagger}_j \widehat{L}_j.
\label{masterdef}
\ee
 If the further 
assumption is made that each Lindblad operator, $\widehat{L}_j$, 
is a linear function of 
$\widehat{p}$ and $\widehat{q}$, we will show that the evolution of 
$\tW_t(\vxi)$ is the product of the classically evolved $\tW_0(\vxi)$ 
with $g_t(\vxi)$, a Gaussian centred on $\vxi = \zero$ that has 
diminishing width. One can then generalize in a straightforward
way the exact solution given by Agarwal \cite{Aga71} for the Wigner 
function. This is a convolution 
of the Fourier transform of $g_t(\vxi)$ with the classically evolved 
$W_0(\x)$. In other words the Wigner function is coarse-grained by
a widening Gaussian window.

A simple example of symplectic evolution of an open quantum system 
is that of a dust particle interacting with air molecules,
or radiation, so that in the absence of gravity, $\hat H=\hat p^2/2m$
and the interaction with the environment depends basically on the particle's
position: $\hat L= \eta \hat q$, where $\eta$ is the coupling parameter.
This example is discussed by Giulini et al. in \cite{Giu:book}.
Another important example is that of an optical field, say an arbitrary
superposition of coherent states interacting with thermal photons. In
terms of real variables, the internal Hamiltonian is just 
$\widehat{H} = \omega ( \widehat{p}^2 + \widehat{q}^2 )/2$,
i.e. the harmonic oscillator. The Lindblad operators in this case
are known to be $\gamma (\bar{n}+1)\widehat{a}/2$ and 
$\gamma\bar{n}\widehat{a}^{\dagger}/2$, where $\widehat{a}^{\dagger}$ and 
$\widehat{a}$ are the usual creation and annihilation operators, $\bar{n}$
is the average number of thermal photons at the frequency $\omega$ of the 
cavity mode at temperature $T$ and $\gamma$ is the decay rate \cite{KimBuz92}.

Recently Di\'osi and Kiefer \cite{DioKie02} (hereafter labeled DK) 
showed in the case of the first example that the Wigner function of any
pure state becomes positive within a definite time. Thus the Markovian
interaction with the environment has the effect of erasing the interference 
fringes characteristic of quantum coherence and from then on the effect of
coarse-graining on the Wigner function is not distinguishable from that of
a classical Liouville distribution.
How general is the DK scenario? In section $II$ we present the exact solution 
of the Lindblad equation for general quadratic $\widehat{H}$ and arbitrary 
 complex linear Lindblad operators 
$\widehat{L}_j = \lambda_j \widehat{q} + \mu_j \widehat{p}$. 
In section $III$ we explicit the underlying classical structure of
the solution.
Then, in section $IV$, we use the properties of the convolution 
to generalize the positivity time of DK 
in the case of arbitrary quadratic Hamiltonian and nonhermitian Lindblad
operators. Furthermore we make the much stronger statement that 
the Wigner function cannot be positive before this threshold, 
unless the initial
distribution is a Gaussian. We discuss the consequences of this
statement through the example of a bath of photons.
In section $V$ we specify the behaviour of the positivity threshold  
for each type of quadratic Hamiltonian. It turns out that  a nonzero 
dissipation coefficient implies that positivity is reached exponentially fast.
The positivity 
threshold is in general inversely proportional to  the 
dissipative coefficient. However it becomes inversely proportional to  the 
Lyapunov exponent if the latter is greater than this coefficient, 
that is in the case of a hyperbolic
system (i.e. the inverted oscillator) in the weak coupling limit. Then 
positivity can be reached much faster  than in the corresponding
elliptic case (i.e. the harmonic oscillator with the same coupling
constants).
Though all the formulae presented here are 
appropriate for a single degree of freedom, the generalization to higher 
dimensions is discussed in this section.
In section $VI$ we  derive a general formula for the growth of the linear 
entropy $(1-\mathop{Tr}{\widehat{\rho_t}^2})$, with respect to the initial 
density operators, 
for each choice of the quadratic master equation. We also show that for
long times the growth of linear entropy attains a universal form.
Finally in section $VII$ we point out that this general solution
is obviously reversible, giving a very synthetic inversion formula
which generalizes previous work about quantum state reconstruction
\cite{KisHer95}. 

The generalization of the convolution  as exact solution of
the master equation (\ref{masterdef}) when $\widehat{H}$ is not quadratic, or 
for nonlinear $\widehat{L}_j$, is not obvious. However, the approximate 
semiclassical theory developed by one of the present authors \cite{Alm02}
has no such constraint. Its compatibility with the present theory is the 
subject of a companion paper \cite{BroAlm03-b}. A simpler version of the 
present work, for the restricted case of hermitian $\opL_j$, can be
accessed in \cite{BroAlm03}.

\section{Exact Solution in the quadratic case with dissipation}
\label{sec-genDK}

 We derive here the exact solution of the Lindblad equation in
the case where the Hamiltonian is quadratic and the Lindblad
operators are complex linear forms in $\oq$ and $\op$. 

Taking the Weyl-Wigner transform of equation (\ref{masterdef}), i.e.
associating the Weyl symbol $A(\x)$ to each operator $\widehat{A}$, as 
defined  in (\ref{wignerdef}), and using the product rules \cite{Gro46}
for operators, we obtain,
\begin{eqnarray}
\disp \frac{\der W_t}{\der t}(\x) & = & 
\disp \left\{ H(\x) , W_t(\x) \right\} + 
\sum_j (\J\vl''_j\cdot\vl'_j)\biggl[\x\cdot\frac{\der W_t}{\der \x}(\x) + 
2 W_t(\x) \biggr] + {}   \nonumber \\
& & {}+ \frac{\hbar}{2} \sum_j
\left\{ \J\vl'_j\cdot\left(\frac{\der^2 W_t}{\der \x^2}(\x)\right) \J\vl'_j
+  \J\vl''_j\cdot\left(\frac{\der^2 W_t}{\der \x^2}(\x)\right) \J\vl''_j 
\right\}.
\label{lind-genDK}
\end{eqnarray} 
Here the $L_j(\x)=\vl'_j\cdot\x+i\vl''_j\cdot\x$ are the Weyl symbols  
of the nonhermitian linear Lindblad operators. We use the notation
\be
\begin{array}{lllllllllll}
\x & = & \left(
\begin{array}{c}
p \cr
q
\end{array} \right) 
& {\textrm and} & \vl' & = & \left(
\begin{array}{c}
\lambda'  \cr
\mu'
\end{array} \right), 
& \vl'' & = & \left(
\begin{array}{c}
\lambda''  \cr
\mu''
\end{array} \right) 
\end{array}.
\label{vldef}
\ee
It is well known that in this case $H(\x)$ and $L_j(\x)$ can be
identified as the classical variables corresponding to
$\widehat{H}$ and $\widehat{L_j}$. Note that if $\vl''=0$ then the second 
term in (\ref{lind-genDK}) cancels. It will become clear that this term
is responsible for dissipation in the evolution of $W_t(\x)$ and we
define the dissipation coefficient, $\alpha=\sum_j (\J\vl''_j\cdot\vl'_j)$,
which is zero in \cite{BroAlm03}.

It is actually easier to solve the evolution equation for the chord function
 $\tW(\vxi)$, 
\be
\frac{\der \tW_t}{\der t}(\vxi) = \left\{ H(\vxi), \tW_t(\vxi) \right\} -
\alpha \vxi\cdot\frac{\der \tW_t}{\der \vxi}(\vxi) - 
\frac{1}{2\hbar} \sum_j \biggl[\left( \vl'_j\cdot\vxi \right)^2 +
\left( \vl'_j\cdot\vxi \right)^2 \biggr] ~ \tW_t(\vxi),
\label{lindchord-genDK}
\ee 
as derived in the appendix A.
We guess a solution of the form
\be
\tW_t(\vxi) = \tW_0(\vxi_{-t})
\exp{ 
  \left( 
        -\frac{1}{2\hbar} \sum_j \int_{0}^{t} 
 \biggl[
\left( \vl'_j\cdot\vxi_{t'-t} \right)^2 + \left( \vl''_j\cdot
\vxi_{t'-t} \right)^2 
\biggr]
 ~ dt' 
  \right)
    },
\label{guess-genDK}
\ee
 where $\vxi_t$ is a 
linear evolution of $\vxi$, which will be explicited a posteriori, such that
\be
\vxi_0 = \vxi.
\ee
 Then, inserting the form (\ref{guess-genDK}) of $\tW_t$ in equation 
(\ref{lindchord-genDK})
and dividing both sides by the exponential of
(\ref{guess-genDK}) leads to the following left part,
\begin{eqnarray}
\lefteqn{ 
-\frac{\der\tW_0}{\der \vxi}(\vxi_{-t})\cdot \dot{\vxi}_{-t}
- \frac{1}{2\hbar}\sum_j\biggl[\left(\vl'_j\cdot\vxi\right)^2
+ \left(\vl''_j\cdot\vxi\right)^2 \biggr] \tW_0(\vxi_{-t})
{} } \nonumber \\
& & {} 
 - \frac{1}{2\hbar}\tW_0(\vxi_{-t})\sum_j 
\int_0^t 
\biggl[ 
2\left( \vl'_j\cdot\vxi_{t'-t} \right)
\vl'_j \cdot \left( - \dot{\vxi}_{t'-t} \right) 
+ 2\left( \vl''_j\cdot\vxi_{t'-t} \right)
\vl''_j\cdot \left( \dot{\vxi}_{t'-t} \right)
\biggr] ~dt',
\end{eqnarray}
which must equal the right part
\begin{eqnarray}
\lefteqn{
-2\J\mH\vxi_{-t}\cdot\frac{\der\tW_0}{\der \vxi}(\vxi_{-t})
-\alpha \vxi_{-t}\cdot\frac{\der \tW_0}{\der \vxi}(\vxi_{-t})
- \frac{1}{2\hbar}\sum_j\biggl[\left(\vl'_j\cdot\vxi\right)^2 + 
\left(\vl''_j\cdot\vxi\right)^2
\biggr]\tW_0(\vxi_{-t})
 - \frac{1}{2\hbar}\tW_0(\vxi_{-t}) \sum_j 
{} } \nonumber \\
& & {} 
\int_0^t
2 \biggl[ \left(\vl'_j\cdot\vxi_{t'-t}\right)
\vl'_j\cdot\left( -2\J\mH\vxi_{t'-t} - \alpha \vxi_{t'-t} \right)
+  \left(\vl''_j\cdot\vxi_{t'-t}\right)
\vl''_j\cdot\left( -2\J\mH\vxi_{t'-t} - \alpha \vxi_{t'-t} \right)
\biggr] ~dt'  .
\end{eqnarray}
We have used
\be
\vxi\cdot\frac{\der}{\der \vxi}\left(\tW_0(\vxi_t)\right) = \vxi_t\cdot 
\frac{\der\tW_0}{\der \vxi} (\vxi_t),
\ee
and 
\be
\left\{ H(\vxi), \tW_t(\vxi) \right\} = 
-2\J\mH\vxi\cdot\frac{\der\tW_t}{\der \vxi}(\vxi).
\ee
 Hence the ansatz (\ref{guess-genDK}) is a solution of (\ref{lindchord-genDK})
 if $\vxi_t$ fulfills
\be
\dot{\vxi}_{t} = 2\J\mH\vxi_{t} + \alpha \vxi_{t}.
\ee
Thus, we can write explicitly,
\be
\vxi_t = e^{\alpha t} \R_{t} \vxi,
\ee
where $\R_t$, defined in (\ref{defRt}), gives the purely Hamiltonian 
evolution and the dissipation term $\alpha$ leads to a classical 
nonhamiltonian  expansion ($\alpha>0$) or contraction ($\alpha<0$) of
the chord variable $\vxi$. One should be aware that although the
Hamiltonian part of the evolution of $\vxi$ is shared with the one
of the phase space point $\x$, the effect of dissipation is inverted, as it
will be explicited soon. 

The argument of the exponential in (\ref{guess-genDK}) is a quadratic 
form in $\vxi$, so the solution can be written
\be
\tW_t(\vxi) = \tW_0(\vxi_{-t})  \exp{\left( 
-\frac{1}{2\hbar} \vxi\cdot\M(t)\vxi
\right)},
\label{exact-sol-chord}
\ee
 with $\M(t)$ a real, time dependent $2\times 2$ matrix, which can naturally
be decomposed into
\be
\M(t) = \sum_j M_j(t) = \sum_j 
\int_0^t dt' e^{2\alpha(t'-t)} \R_{t'-t}^{T} \vl_j \vl_j^{T} \R_{t'-t},
\label{defM}
\ee
so that each Lindblad operator contributes a Gaussian to 
(\ref{exact-sol-chord}).

 Now, back into the Weyl-Wigner representation by using (\ref{chordtrans}), 
one obtains the solution of (\ref{lind-genDK}),
\be
\W_t \left( \x \right) = \frac{1}{2\pi\hbar} e^{2\alpha t}
\int W_0 \left( e^{\alpha t}\R_{-t}(\x - \y) \right)
 \frac{1}{\sqrt{\det \M_J(t)}} \exp{\left( 
- \disp \frac{1}{2\hbar} \y\cdot\M_J(t)^{-1}\y
\right)}~d\y,
\label{exact-sol-weyl}
\ee
where we have defined
\begin{eqnarray}
\M_J(t) & = & -\J \M(t) \J \cr
\M_J(t)^{-1} & = & -\J \M(t)^{-1}\J,
\end{eqnarray}
with the symplectic matrix $\J$ defined in (\ref{wedgedef}).
$W_0(\x)$ is the initial Wigner function and the convolution Gaussian turns
into a Dirac $\delta$ function as $t$ goes to $0$.
We have equivalently,
\begin{eqnarray}
W_t \left( \x \right) & = & w_t(\x_{-t}) \\
~ & = &
\frac{1}{2\pi\hbar\sqrt{\det \M_J(t)}} 
\int W_0 \left( \y \right)
\exp{\left( 
- \disp \frac{1}{2\hbar} 
(\y-\x_{-t})\cdot\widetilde{\M}(t)^{-1}(\y-\x_{-t})
\right)} ~d\y, 
\label{exact-sol-weyl_2}
\end{eqnarray}
with 
\be
\x_t = e^{-\alpha t}\R_{t}\x
\label{defxt}
\ee
and
\be
\widetilde{\M}(t) = - e^{2\alpha t}\R_{-t}^T\J M(t) \J\R_{-t} = -\M_J(-t).
\ee
Hence the solution is a convolution with a Gaussian
which broadens in time, composed with a backwards 
nonhamiltonian evolution of the phase space variable $\x$.
As mentioned earlier, the Hamiltonian part of this classical evolution
is the same as in the chord space, whereas the dissipative part has the
opposite effect: dissipation ($\alpha>0$) will shrink the phase space
variable, and thus expand the distribution $W_t$.

 This solution, which has been derived in the case of an homogeneous 
 quadratic Hamiltonian, can be generalized easily to a quadratic Hamiltonian
with a linear part. One then has to be aware that the matrix $\R_t$, 
which appears in the exponential Lindbladian damping, strictly corresponds to 
the classical motion of the {\it chord}, which is determined by the 
{\it homogeneous} part of the Hamiltonian $H$. This remark is important,
for instance, in the parabolic case, say a particle with a linear
potential, where the motion of the chord disregards the potential.

  Since $|\det{\widetilde{\M}(t)}|$ grows with time, one can conclude, 
following DK \cite{DioKie02},
that the solution (\ref{exact-sol-weyl}) becomes positive after a 
certain time. Indeed it is, after rescaling the variable, 
the convolution of 
the initial Wigner function with a Gaussian of broadening 
size, which smoothes out oscillations around zero. It is the 
property of symplectic invariance of the Wigner function that DK employ to 
prove strict positivity in a specific case that is now extended to its 
broader context. Moreover we shall give in section $IV$ the much stronger 
result that  the Wigner function cannot be positive before
the DK time, which does not depend on the initial pure state, 
unless it is a coherent state. 
We then give in section $V$  the general behaviour of this 
positivity threshold for different dynamics, namely when $\mH$ is elliptic, 
as for the  harmonic oscillator, 
 hyperbolic, as for the scattered particle, or in the parabolic intermediate 
case which includes the system studied in DK.
The next section explicits the formal correspondence between
this problem and a classical Brownian motion described by a Langevin 
equation, as also did Agarwal \cite{Aga71} for his solution.

\section{Classical correspondence}

 Since Eq. (\ref{lind-genDK}) is a Fokker-Planck equation, it can be 
interpreted
as the evolution equation for the probability distribution of a classical
Brownian motion defined by a Langevin equation. This correspondence
gives a simple classical interpretation of the problem. Hence the
decoherence may be seen as a diffusion induced by some random force, 
and dissipation can be interpreted as a classical viscosity, although 
it is always  accompanied by another diffusive term. The only feature
which cannot be assigned a classical meaning is the Wigner function
itself, which, as a  pseudo-probability distribution, can have negative
values.

One can check easily, see for instance \cite{Str63}, that the following 
Langevin equation, 
\be
\left\{
\begin{array}{lll}
\dot{p} & = & -\frac{\der H}{\der q}(\x) - \alpha p + \sqrt{\hbar}
\sum_m \left( \lambda'_m f_m(t) + \lambda''_m g_m(t) \right)  \cr
\dot{q} & = & \frac{\der H}{\der p}(\x) - \alpha q  + \sqrt{\hbar}
\sum_m \left( \mu'_m f_m(t) + \mu''_m g_m(t) \right)
\end{array}
\right.,
\ee
induces equation (\ref{lind-genDK}) as a Fokker-Planck counterpart.
The ``Brownian forces'' $f_m(t)$ and $g_m(t)$ verify
\be
\begin{array}{lll}
\langle f_m(t') f_n(t'') \rangle & = & \delta_{m,n} \delta(t) \cr
\langle g_m(t') g_n(t'') \rangle & = & \delta_{m,n} \delta(t) \cr
\langle f_m(t') g_n(t'') \rangle & = & 0 \cr
\end{array}.
\label{Langevin1}
\ee
The $f_m(t)$ correspond to the diffusion induced by the 
nondissipative real part of the Lindblad operators, whereas the $g_m(t)$ 
correspond to the diffusion induced by dissipation.
It can easily be verified that the Fokker-Planck equation is symplectically
invariant, so one is allowed
to perform the following change of coordinates,
\be
\left\{
\begin{array}{lll}
\bar{p}  & = &  p - \frac{\alpha}{2\mH_{11}} q\cr
\bar{q} & = & q
\end{array}
\right.,
\ee
to the above Langevin equation, which then turns into the following 
more intuitive form,
where the dissipation depends only on the momentum:
\be
\left\{
\begin{array}{lll}
\dot{p} & = & -\frac{\der \bar{H}}{\der q}(\x) - \bar{\alpha} p + \sqrt{\hbar}
\sum_m \left( \bar{\lambda}'_m f_m(t) + \bar{\lambda}''_m g_m(t) \right) \cr
\dot{q} & = & \frac{\der \bar{H}}{\der p}(\x) + \sqrt{\hbar}
\sum_m \left( \mu'_m f_m(t) + \mu''_m g_m(t) \right)
\end{array}
\right..
\label{Langevin2}
\ee
Here, the matrix for the transformed Hamiltonian $\bar{H}$ is
\be
\left(
\begin{array}{cc}
\mH_{11} & \mH_{12} \cr
\mH_{12} & \mH_{22}+\frac{\alpha^2+4\alpha \mH_{12}}{4\mH_{11}}
\end{array}
\right),
\ee
$\bar{\lambda}'_m = \lambda'_m - \frac{\alpha}{2\mH_{11}} \mu'_m$ and
respectively for $\bar{\lambda}''_m$, and $\bar{\alpha} = 2\alpha$.

From this classical picture we can interpret the general behaviour of
the solution of the Lindblad equation. 
In the case of a closed system, remember that the Wigner function
undergoes a Liouville, unitary, propagation. 
Now the system is coupled to an environment, i.e. there are nonzero
 Lindblad operators.
If dissipation of energy is neglected, these operators are Hermitian,
so there is no imaginary part of the vectors $\vl$.
Then the effect of the environment over the system can be interpreted as
a diffusion process corresponding to random forces in the Langevin equation. 
Formally, it corresponds to the initial Wigner function
being convoluted with a Gaussian which broadens with time.
If one now takes into account the dissipation induced by the environment, 
allowing the Lindblad operator to be non-hermitian, a viscous term appears
in (\ref{Langevin1}) and (\ref{Langevin2}), meaning that the classical
trajectories on which the distribution travels are drifted to lesser
or higher energy, according to the sign, $+$ or $-$, of the ``viscosity''
$\alpha$. Indeed, the dissipative linear motion governed by the 
nonrandom terms of (\ref{Langevin1}) is just that of (\ref{defxt}).
One should be aware that it is only a formal classical scheme, and that
the viscous term might have a purely quantum origin.
For instance, in the case of photons in a cavity with dissipation, 
whose Lindblad operators are explicited
in section $V-A$, the viscosity is related to spontaneous emission, which
breaks the symmetry between emission and absorption. Then the classical 
trajectories of the above equation spiral in towards the origin, although
a semiclassical theory would lead to no dissipation. 
The opposite case would be an amplified cavity, where the trajectories
would spiral out.
Notice that this viscous term always goes along with a supplement of
diffusion, which can be interpreted as a consequence of the 
fluctuation-dissipation ``theorem'' \cite{MeySar:book}. 

\section{Unique positivity time for any initial state}

 We have seen in section $II$ that, because of the Lindbladian
part of the master equation, the pure state Wigner function is
convoluted with a Gaussian whose width grows with time. 
It has been pointed out by DK\cite{DioKie02} that at the time $t_p$
at which the width of the Gaussian reaches $\hbar$,
that is when it becomes the Weyl representation of some coherent or 
squeezed state,
then its convolution with the initial Wigner function, $W_0(\x)$, is a
Husimi function \cite{Hus40}\cite{Vor77}\cite{Alm98} 
of the initial state, or a $Q$ function in the language 
of quantum optics. It is a well known property
that the Husimi function is positive, so we have $W_{t_p}\geq 0$, and
obviously, since the Gaussian is strictly broadening, $W_{t}> 0$ for $t>t_p$.
It has already been emphasized by Leonhardt et al. \cite{LeoPau93}
 that the form of the Wigner function
after interaction with a dissipative environment
 can be read as an 
intermediate phase space distribution $W(\x,t=0,s)$, with an $s$ which
depends on the dissipation rate. Thus $s=0$ corresponds
to the (initial) Wigner function and $s=-1$ to the Husimi function of DK
(in \cite{LeoPau93} the role of the environment is played by
the imperfections of a beam separator).

We shall now  prove that 
the Wigner function can never be positive before the
positivity threshold $t_p$, unless it is positive from the beginning, 
that is, unless the initial state is a Gaussian state.
Indeed, if an initial pure state $|\psi_0\rangle$
is not a Gaussian, then it was shown by Tatarski\u{\i}
\cite{Tat83} that the initial 
Wigner function $W_0(\x)$ has negative parts. But it is also true that 
nongaussian Husimi functions necessarily have zeroes, 
as shown in the appendix B.  Since $W_{t_p}$ is, up to 
a symplectic transform, a Husimi function, there exists $\x_0$ such that
\be
W_{t_p}(\x_0)=0.
\ee
Let us now investigate the case $t<t_p$. 
Then $W_t(\x)$ given by (\ref{exact-sol-weyl_2}) is a 
convolution of $W_0(\x_{-t})$
with a Gaussian of width smaller than $\hbar$.
The point is that one can then convolute again 
with $\exp{\left(-\frac{\kappa}{\hbar}\x_{-t}\cdot
\widetilde{\M}(t)^{-1} \x_{-t} \right)}$, with the real parameter 
$\kappa$ chosen so that the output is also a Husimi 
function $Q(\x_{-t})$. 
To show this we refer to the simple general relation:
\be
\int d\y~\exp{\left(-(\x-\y)\cdot\M(\x-\y)\right)} 
\exp{\left(-\y\cdot\kappa \M\y\right)} 
= \frac{\pi}{(1+\kappa)\sqrt{\det{\M}}}
\exp{\left(-\x\cdot\frac{\kappa}{1+\kappa} \M\x\right)}.
\ee
This Husimi function can be identified with $W_{t_p}$ through a symplectic 
transform $\x_{-t}\rightarrow\x'$, so we have
\be
Q(\x'_0) = \int d\y~ w_t(\x'_0-\y) 
\exp{\left(-\frac{\kappa}{2\hbar}\y\cdot\widetilde{\M}(t)^{-1}\y\right)} = 0.
\label{convol-tinftp}
\ee
Now it is obvious from (\ref{convol-tinftp}) that $w_{t}$, hence $W_{t}$, must 
have a negative part for all $t<t_p$.

Let us emphasize the striking consequence of this result: the positivity
time does not depend on the initial distribution, as long as it is
not a Gaussian one. The following example shall illustrate this remarkable
property in a more transparent way.

\subsection{Example}
We start from the familiar superposition of two coherent states, 
i.e. ground states of the harmonic oscillator, displaced to the
phase space points $\Zeta = (0,\pm\zeta)$:
$|\psi_0\rangle = (|\zeta\rangle+|-\zeta\rangle )/\sqrt(2)$, in
the context of photons in a cavity with dissipation. The Wigner function,
which is a particular case of Eq. (\ref{exact-sol-weyl}),
is the sum of three terms; two of these correspond to the coherent states
taken independently, and the third term comes from their  interference:
\be
W_t(\x) = W_{\zeta}(\x) + W_{-\zeta}(\x) + W_{i}(\x),
\ee
with
\be
W_{\zeta}(\x) = \frac{2N}{\pi \beta_t}\exp{\left(-\frac{2}{\beta_t}p^2\right)}
\exp{\left(-\frac{2}{\beta_t}(q-e^{-\gamma t/2}\zeta)^2\right) }
\ee
and
\be
W_{i}(\x) = \frac{4N}{\pi \beta_t}\exp{\left(-\frac{2}{\beta_t}(p^2+q^2)\right)}
\exp{\left(-2(1-\frac{e^{-\gamma t}}{\beta_t})\zeta^2\right)}
\cos{\left(\frac{4e^{\gamma t/2}}{\beta_t}\zeta p\right)},
\ee
where $\gamma$, defined in the introduction, corresponds to $2\alpha$,
and $\beta_t = 2 \bar{n}(1-\exp{\left(-\gamma t\right)})+1$. The function 
is normalised by $N=\left(1+\exp{(-\zeta^2/\hbar)}\right)^{-1}$.
Obviously the minimum values are concentrated on the line $q=0$, where 
the expression simplifies into
\be
W_{t}(p,q=0) = \frac{4N}{\pi \beta_t}\exp{\left(-\frac{2}{\beta_t}p^2\right)}
\Biggl[
\exp{\left(-2(1-\frac{e^{-\gamma t}}{\beta_t})\zeta^2\right)}
\cos{\left(\frac{4e^{\gamma t/2}}{\beta_t}\zeta p\right)} + 
\exp{\left(-2\frac{e^{-\gamma t}}{\beta_t}\zeta^2\right) }
\Biggr].
\ee
The Wigner function becomes positive when 
\be
1-\frac{e^{-\gamma t}}{\beta_t} = \frac{e^{-\gamma t}}{\beta_t},
\ee
that is at $t_p=1/\gamma\log{\left(1+1/(2\bar{n}+1)\right)}$, which
indeed does not depend on the initial spacing $\zeta$ of the two
coherent states. On the other hand, the position $(p_m,0)$ of the closest 
zero, given by the first minimum of the cosine at that time,
\be
p_m = \frac{\beta_t}{4\sqrt{2}\zeta},
\ee
gets further away as $\zeta$ becomes smaller. This shows that
the negative regions that remain until $t_p$ may be so shallow as to be 
practically irrelevant.

The chord representation reveals how the positivity threshold is
related to a more reasonable estimate of the decoherence time.
The expression of the chord function in our example is
\be
\tW_0(\vxi) = \frac{1}{2\pi\hbar}
\exp{\left(-\frac{\vxi^2}{4\hbar}\right)}
2\cos{\left(\frac{\vxi\wedge\Zeta}{\hbar}\right)} 
+ \frac{1}{2\pi\hbar} \exp{\left(-\frac{(\vxi-\Zeta)^2}{4\hbar}\right)}
+ \frac{1}{2\pi\hbar} \exp{\left(-\frac{(\vxi+\Zeta)^2}{4\hbar}\right)}.
\ee
The first two terms, distributed around the origin, correspond
to the two coherent states taken separately, whereas the last two terms,
distributed around $\Zeta$ and $-\Zeta$, describe the quantum interference
between them.
The positivity time corresponds to the moment when the 
Gaussian in (\ref{exact-sol-chord}) damps everything
outside a region of size $\hbar$ in the chord space.
However, both interference terms in this example will be
damped much sooner if $\zeta$ is large enough, indicating an overall loss of
coherence.

It should be remarked that one just has to study
a specific example, here the twin coherent states, to get the
positivity time for any nongaussian initial pure state.

\section{Behaviour of the positivity threshold}

Positivity is attained when the determinant of the matrix 
$\widetilde{\M}(t)$, that
is the determinant of $\M(-t)$, is equal to $1/4$. Then the expression
of the solution (\ref{exact-sol-weyl_2}) is indeed a Husimi function.
The matrix $\M$ is defined by (\ref{defM}), so, to calculate it, we 
 notice that $\J\mH $, in the expression (\ref{defRt}) of $\R_t$, 
can be diagonalized in most cases, that is when its two 
eigenvalues are finite and different. We then define the 
matrix $\mP$ such that
\be
2\J\mH = \mP^{-1}\mD\mP,
\ee
with
\be
\mD = \left( \begin{array}{cc}
\sigma & 0 \\
0 & - \sigma \end{array} \right),
\label{defD}
\ee
and $\sigma = 2 \sqrt{-\det{\mH}}$.
Notice that since $\mH$ is symmetric, then $\J\mH$ has a null trace, 
and so has $\mD$. The dissipation parameter, $\alpha$, and $\sigma$
are basic elements for the description of the evolution of 
Markovian quadratic open systems.

Then, using (\ref{defM}) one can easily derive the expression of $\M$,
\be
\M(t) = \mP^{T}\left( \begin{array}{cc}
\frac{ 1 - e^{-2(\sigma+\alpha)t} }{2(\alpha+\sigma)}\mA_{11} 
& \frac{ 1 - e^{-2\alpha t} }{ 2\alpha }\mA_{12} \\
\frac{ 1 - e^{-2\alpha t} }{ 2\alpha }\mA_{21} 
& \frac{ 1 - e^{2(\sigma-\alpha)t} }{2(\alpha-\sigma)}\mA_{22} 
\end{array} \right) \mP,
\label{defMt}
\ee
where $\mA$ is defined by
\be
\mA =  (\mP^{-1})^T \sum_j 
\left( \vl'_j (\vl'_j)^T + \vl''_j (\vl''_j)^T \right) \mP^{-1}.
\ee

The corresponding quadratic form actually depends on the classical motion
of the chord, determined by the homogeneous part $\x\cdot\mH\x$ 
of the Hamiltonian. One then has to separate different cases: 
the elliptic case, $\det{\mH} > 0$, the parabolic case, $\det{\mH} = 0$, and 
the hyperbolic case, $\det{\mH} < 0$.

Since the Wigner function is symplectically invariant, one just has to 
treat the simplest expression in each case, respectively the harmonic
oscillator, $H(\x) = p^2/2 + q^2/2$, the particle in a linear
potential, $H(\x) = p^2/2 + q$, and the scattered particle, $H(\x) = pq$,
which is symplectically equivalent to $p^2/2-q^2/2$
\footnote{The reduction of a linear Lindblad operator under a symplectic 
transformation, $\x'=\C\x$, is especially simple in the Wigner or in the 
chord representation. If, instead of (\ref{vldef}), we set 
$L_j(\x) = \l_j\wedge\x$, then the invariance is obtained by taking $\l_j'=\C\l_j$. 
Of course, one must also use the invariance of the classical Hamiltonian 
$H(\x')=H(\x)$.}.

\subsection{Harmonic oscillator}
In this case, the Hamiltonian reads
\be
H(\x) = \omega\left( \frac{p^2}{2} + \frac{q^2}{2} \right)
\ee
and the matrix $\mP$ is 
\be
\mP = \frac{1}{\sqrt{2}} \left( \begin{array}{cc}
-1 & i \\
-i & 1 
\end{array} \right).
\ee
Then the determinant of $\M(-t)$ reads
\be
\det{\M(-t)} = 
\frac{e^{4\alpha t} -2e^{2\alpha t}\cos{2\omega t} +1}{4(\alpha^2+\omega^2)}
\mA_{11}\mA_{22} - \frac{e^{4\alpha t} - 2e^{2\alpha t} + 1}{4\alpha^2}
\mA_{12}\mA_{21} ,
\ee
where in this case the matrix $\mA$ is complex:
\be
\mA = \frac{1}{2} \sum_j \left( \begin{array}{cc}
(\lambda'_j)^2 - (\mu'_j)^2 + 2i\lambda_j\mu_j  &
-i (\lambda'_j)^2 - i (\mu'_j)^2 \\
 -i (\lambda'_j)^2 - i (\mu'_j)^2 &
- (\lambda'_j)^2 + (\mu'_j)^2 + 2i\lambda_j\mu_j \end{array} \right)
+ \left(\textrm{idem with} ~\lambda''_j ~\textrm{and} ~\mu''_j \right),
\ee
using (\ref{vldef}).

In the dissipative case, that is for $\alpha>0$,  this determinant diverges 
exponentially fast, and 
positivity is attained in a time of the order of $1/|\alpha|$.  
Let us take for instance the bath of 
photons, then the coefficients of the Lindblad operators are
\be
\begin{array}{cccccccccccc}
\vl'_1 & = & \left( \begin{array}{c} 0 \\ \sqrt{\frac{\gamma(\bar{n}+1)}{2}} 
\end{array} \right), & 
\vl''_1 & = & \left( \begin{array}{c} \sqrt{\frac{\gamma(\bar{n}+1)}{2}} 
\\ 0 \end{array}  \right), &
\vl'_2 & = & \left( \begin{array}{c} 0 \\ \sqrt{\frac{\gamma\bar{n}}{2}} 
\end{array} \right), &
\vl''_2 & = & \left( \begin{array}{c} - \sqrt{\frac{\gamma\bar{n}}{2}} 
\\ 0 \end{array} \right),
\end{array}
\ee
and the friction $\alpha = \gamma/2$.
Then 
\be
\det{\M(-t)} = \frac{ ( e^{\gamma t} - 1 )^2 }{4}(2\bar{n}+1)^2,
\ee
which equals $1/4$ at $t = 1/\gamma\log{\left(1+1/(2\bar{n}+1)\right)}$, as
was previously mentioned.
Notice that, although the coarse-graining grows for ever, the size of 
the Wigner distribution
reaches a finite limit, for the rescaled function
(\ref{exact-sol-weyl_2}) is the expression of $W(\x)$ and
not $W(\x_{-t})$.

If, on the other hand, $\alpha<0$ then the determinant reaches its limit in a 
time also of the order of $1/|\alpha|$. We conjecture that
this limit has a lower bound greater than $1/4$.
Here rescaling $\x_{-t}\rightarrow\x$ 
now implies that $W_t(\x)$ spreads with no bound.

\subsection{Scattered particle}
The simplest form of the Hamiltonian for a particle scattered by a parabolic
barrier is
\be
H(\x) = \omega pq.
\ee
Then the matrix $\mP$ is just identity and the matrix $\mA$ is real, 
so the determinant reads
\be
\det{\M(-t)} = 
\frac{e^{4\alpha t} - 2e^{2\alpha t}\ch{2\omega t} + 1}{4(\alpha^2-\omega^2)}
\mA_{11}\mA_{22} - \frac{e^{4\alpha t} - 2e^{2\alpha t} + 1}{4\alpha^2}
\mA_{12}\mA_{21} ,
\ee
As long as $\alpha>-\omega$ it grows exponentially, so positivity is
always reached. 
When $\alpha<-\omega$ the determinant has again a finite asymptotic value.
Then the positivity threshold is of the order of $1/(|\alpha|-\omega)$,
which is greater than the corresponding elliptic case, with identical
Lindblad operators.

The main difference with the elliptic case appears in the weak coupling
limit $|\alpha|\ll\omega$. Whereas positivity of the elliptic system 
will then be attained in a time which is still inversely proportional to
the coupling with the environment, the positivity threshold of the 
hyperbolic system will saturate at $1/\omega$. We conjecture
that this will also be the case in a more general chaotic system.

\subsection{Particle in a linear potential}
We now study the intermediate case, which can be represented by the Hamiltonian
\be
H(\x) = \frac{p^2}{2} + q.
\ee
This degenerate case does not follow our general form for the matrix $\M$,
so one has to treat it separately, taking care of the linear term (cf. remark
of section $II$).
One should remember here that the motion of the chord is given by the 
free motion of the particle, which corresponds to
\be
\R_t = \left(\begin{array}{cc}
1 & 0 \cr
t & 1
\end{array}\right).
\ee
Then the damping matrix is
\be
\M(t) = t \sum_j \left(\begin{array}{cc}
\disp  \frac{e^{-2\alpha t}}{2\alpha}
Q^{(2)}_{11}  
-\frac{1}{2\alpha} Q^{(0)}_{11} & 
\disp  \frac{e^{-2\alpha t}}{2\alpha}
Q^{(1)}_{12} 
-\frac{1}{2\alpha} Q^{(0)}_{12} \cr
\disp  \frac{e^{-2\alpha t}}{2\alpha}
Q^{(1)}_{12} 
-\frac{1}{2\alpha} Q^{(0)}_{12} & 
\disp \left(\frac{e^{2-\alpha t}}{2\alpha}-\frac{1}{2\alpha}\right)Q^{(0)}_{22}
\end{array}\right),
\ee
where the $Q^{(d)}_{ij}$ are polynomials of degree $d$ in $t$ and of degree $2$
in the coupling constants, say $(\lambda'_j,\lambda''_j,\mu'_j,\mu''_j)$.
Let us take for instance one Lindblad operator with $(\vl')^T = (0,\sqrt{D'})$
and $(\vl'')^T = (\epsilon\sqrt{D''},0)$, with $\epsilon=\pm 1$. Then one has
\be
\det{\M(-t)} = \frac{1}{4} \Bigl[ 
e^{4\epsilon\bar{D}t} \left( 1 + \frac{1}{4(D'')^2} \right)
- e^{2\epsilon\bar{D}t} \left( \frac{D'}{D''}t^2+\frac{1}{2(D'')^2} +2 \right)
+ 1 + \frac{1}{4(D'')^2} \Bigr],
\ee
with $\bar{D}=\sqrt{D'D''}$. The limit is always greater than $1/4$, with
the usual exponential contrast between the dissipative case ($\epsilon=1$)
 and the excited case ($\epsilon=-1$).
In \cite{DioKie02}, DK study this example with no dissipation, and they 
find a positivity time of the order of $1/\sqrt{D'}$. On the following 
table, one can read
different values of the positivity threshold for $D'=2$, and check that
the limit $D''\rightarrow 0$ is attained gradually:
\be 
\begin{tabular}{|c|c|c|c|c|c|}
$D''$ & $0$ & $0.1$ & $1$ & $10$ & $100$ \\
\hline
$\epsilon=-1$ & 0.930 & 0.640 & 0.244 & 0.077 & 0.022 \\
$\epsilon=1$ & 0.930 & 1.040 & 1.025 & 0.752 & 0.400 \\
\end{tabular}.
\ee
However for large values of $D''$, the positivity threshold will behave like
$1/\sqrt{D'D''}$.

So far the discussion has been restricted to the case of a single degree of 
freedom. Besides trivial changes of factors of $2\pi\hbar$, the basic form 
of the solutions of the Lindblad equation in the chord representation 
(\ref{exact-sol-chord}) and for the Wigner function 
(\ref{exact-sol-weyl}) remain unchanged in the case of $n$ degrees 
of freedom. The evolution matrix $\R_t$ now has the dimension 
$(2n)\times(2n)$, but it can again be simplified by symplectic 
transformations. It will often decompose into blocks corresponding to the 
above examples. If every Lindblad vector $\vl_j$ is defined for a single 
block, then its contribution to $\M(t)$ will be of the same form as that for 
a single degree of freedom, but otherwise each case must be examined 
separately. Also, $4-$dimensional blocks may arise, corresponding to 
hyperbolic spiral motion, as well as singular cases analyzed by 
Arnold \cite{Arn:book}.

\section{The growth of linear entropy}

Besides considering the positivity of the Wigner function, we can use the 
exact solution  (\ref{exact-sol-chord}) to investigate the growth of 
linear entropy
\be
S_t = 1- \tr{\widehat{\rho_t}^2}.
\ee
This is only zero for a pure state, just as for the Von Neumann
entropy 
\be
{\mathbb S}_t = -\tr{\widehat{\rho_t}\ln\widehat{\rho_t}}.
\ee
Since the solution is simpler in the chord representation, we make use 
of the following relations\footnote{Note that there is a misprint in 
formula (6.24) of \cite{Alm98}} for operators 
$\widehat{A}$, $\widehat{B}$,\ldots 
represented by chord functions $\widetilde{A}(\vxi)$, 
$\widetilde{B}(\vxi)$ \ldots
\be
\A \rightarrow \tA(\vxi), 
\ee
\be
\A^{\dagger} \rightarrow \tA(-\vxi)^*, 
\ee
\be
\tr{\A} = \tA(\vxi=0), 
\ee
\be
\tr{\A\B} = \frac{1}{2\pi\hbar}\int d\vxi \tA(\vxi)\tB(-\vxi).
\ee
Therefore, using ${\oprho}^{\dagger} = \oprho$ and $\trho_t(\vxi) = 
2\pi\hbar ~ \tW_t(\vxi)$, we obtain
\be
\tr{\oprhot} = 2\pi\hbar ~ \tW_t(\vxi=0) = 1,
\ee
and
\be
\tr{{\oprhot}^2} = 2\pi\hbar \int d\vxi ~ |\tW_t(\vxi)|^2.
\label{trrho2}
\ee
In the case of the solution (\ref{exact-sol-chord}) 
of the Lindblad equation, we thus have 
\be
\tW_0(\vxi=0) = \frac{1}{2\pi\hbar}
\label{tildeW0of0}
\ee
and
\be
2\pi\hbar ~ \int d\vxi ~ |\tW_0(\vxi)|^2 = 1
\ee
if the initial state is pure.

After changing the variables of integration in (\ref{trrho2}), 
replacing $\vxi_{-t}$ in (\ref{exact-sol-chord}) 
by the initial chord $\vxi'$, we obtain
\be
\tr{{\oprhot}^2} = 2\pi\hbar ~e^{2\alpha t} ~ \int d\vxi' ~ |\tW_0(\vxi')|^2 
\exp{\left(\frac{1}{\hbar}\vxi'\cdot\M(-t)\vxi'\right)},
\label{entropy-gen-2}
\ee
where $\M(-t)$, defined in (\ref{defM}), is negative definite.

Thus $\tr{{\oprhot}^2}$ is just the average of a rescaled 
Gaussian whose width 
narrows in time in a manner that depends exclusively on the particular 
form of $H(\x)$ and the linear Lindblad operators $L_j(\x)$.
The initial state merely determines the probability density employed in 
the calculation of the average. 
A general asymptotic behaviour can be predicted for this formula, because 
the width of the Gaussian generally shrinks as $t$ grows. 
If the contraction is
sufficient, the expression (\ref{entropy-gen-2}) will tend to
\be
\tr{{\oprhot}^2} \simeq 2\pi\hbar ~e^{2\alpha t}~|\tW_0(\zero)|^2 \int d\vxi ~ 
\exp{ \Biggl[ \frac{1}{\hbar} \vxi\cdot\M(-t)\vxi \Biggr] } = 
\frac{\pi\hbar e^{2\alpha t}}{ \sqrt{ \det{\M(-t)} } }, 
\label{trrho2asympt}
\ee
by using (\ref{tildeW0of0}). It can be explicited, using (\ref{defD})
and (\ref{defMt}), in terms of the dissipation, $\alpha$, and the basic
Hamiltonian parameter, $\sigma$:
\be
\tr{{\oprhot}^2} \simeq \frac{\pi\hbar }{ \sqrt{ 
\frac{e^{-4\alpha t} - 
2e^{-2\alpha t} \left( \frac{e^{2\sigma t}+e^{-2\sigma t}}{2} \right) 
+ 1}{4(\alpha^2+\sigma^2)}
\mA_{11}\mA_{22} - \frac{e^{-4\alpha t} - 2e^{-2\alpha t} + 1}{4\alpha^2}
\mA_{12}\mA_{21} 
 } }.
\ee

If the underlying classical system is elliptic, 
$\Re{\left(\sigma\right)} = 0$, one can distinguish two situations. In
the excited case, $\alpha<0$, $\tr{{\oprhot}^2}$ converges to zero.
In the dissipative case, $\alpha>0$, 
it converges to a finite value, which is not surprising  since the system then 
reaches a thermal equilibrium \cite{Aga71}. 
For instance in the case of a bath of photons, one has
\be
\tr{{\oprhot}^2} \rightarrow \frac{4\pi\hbar}{2\bar{n}+1}.
\ee
If the system is hyperbolic, the ``Lyapunov exponent''  
$\omega = \Re{\left(\sigma\right)} \neq 0$.
The consequence is to shift the definition of the above dichotomy.
Indeed,  $\tr{{\oprhot}^2}$ has a nonzero limit in the more restricted 
range $\alpha>\omega$. 

The decoherence time, for the decay of $\tr{{\oprhot}^2}$,
is in general inversely proportional to the coupling strength.
However in the weak coupling limit $|\alpha|\ll\omega$ it is $1/\omega$
in the hyperbolic case, whereas it is $1/|\alpha|$ in the elliptic one. 
Hence the decoherence time defined by the linear 
entropy is here consistent with the positivity threshold. This is
a strong support to the thesis that positive Lyapunov exponents
accelerate decoherence \cite{ZurPaz95}.

 The asymptotic formula (\ref{trrho2asympt}) holds only for those cases where 
 all the 
eigenvalues of $\M(-t)$ tend to infinity. 
Counterexamples are the elliptic case with $\alpha<0$, since 
the determinant then has
a finite limit, and the hyperbolic case, 
for $\alpha<-\omega$. Moreover, though $-\omega<\alpha<\omega$ leads to a 
determinant
which tends to infinity, one of the eigenvalues will indeed diverge whereas 
the other one,
corresponding to the unstable direction, will have a finite limit.

\section{Reversibility of the solution}
Although decoherence is usually associated with a loss of information,
which goes along with the growth of entropy, several techniques have
been developed in quantum optics to recover the initial information after 
interaction of the system with the environment. In general these consist of
reconstructing the quantum state of a lossy cavity by using mathematical
inversion formulas \cite{KisHer95} or directly by experimental processes
\cite{MoyRov99}.
We show here that the reversibility of the Wigner function results from 
the deconvolution of its evolution (\ref{exact-sol-weyl_2}), 
or, even simpler,
as a division (\ref{exact-sol-chord}) in the chord space. Indeed, one has
\be
\tW_t(\vxi)=\tW_0(e^{-\alpha t}\R_{-t}\vxi)\tG_t(\vxi),
\ee
which can easily be inverted as
\be
\tW_0(\vxi)=\frac{\tW_t(e^{\alpha t}\R_t\vxi)}
{\tG_t(e^{\alpha t}\R_t\vxi)}.
\ee
This is a generalization and a simplification of the inversion formula 
of \cite{KisHer95} since the loss induced by beam splitter
is a particular case of our general formalism. 

\section{Conclusion}

The exact solution of the Markovian master equation for quadratic 
Hamiltonians and linear complex Lindblad operators has been derived 
in the form of 
a convolution for the Wigner function. This involves the classical evolution 
of the initial Wigner function for the closed system with a phase space 
Gaussian that is independent of this initial condition, while its width 
expands in time, depending only on the Hamiltonian and the Lindblad 
operators. This simple solution  allows for the generalization of 
Di\'osi and Kiefer's proof that the Wigner function becomes positive 
within a definite time $t_p$. Furthermore we support the much stronger 
statement that, unless the initial distribution is already 
a coherent state,
 the Wigner function must have negative regions before that time $t_p$, 
which then 
does not depend on the initial state. In section $V$ we have given the
behaviour of $t_p$ for 
three basic types of motion of the quadratic Hamiltonian, namely the
elliptic, the hyperbolic and the parabolic case, verifying 
that positivity is always reached exponentially fast. The
positivity threshold is generally of the order of $1/|\alpha|$,
except in the weak coupling regime $|\alpha|\ll\omega$ 
of a hyperbolic system, where it is of the order of $1/\omega$.
One should note that the
threshold becomes independent on the Planck constant, if the Lindblad 
equation is appropriately scaled.

The Fourier transform of the exact solution, the chord function, is even 
simpler. This is the product of two terms: one is just the nonhamiltonian
 classical evolution  of the initial chord 
function with dissipation and the other is again a Gaussian, 
but with narrowing width. This leads to a simple formula for $\tr{\oprho^2}$ 
as an average of a shrinking Gaussian, where the probability 
distribution 
used to calculate the mean is just the square modulus of the initial chord 
function. For long times, we use the normalization condition that the chord 
function is unity at the origin to derive simple rules for the asymptotic 
growth $\tr{\oprho^2}$ for a general quadratic Hamiltonian. 
This decays exponentially fast in a time which is of the same order 
as the positivity threshold.
 This result is compatible with the 
arguments presented by Zurek and Paz \cite{ZurPaz95}
for exponential growth of linear entropy for chaotic systems. These are 
classically characterized by local hyperbolicity, where the Lyapunov exponent 
describes the average effected by a typical orbit that approaches 
many hyperbolic points. In contrast, the hyperbolic 
quadratic Hamiltonian defines a linear classical motion, but both will 
exponentially stretch the Wigner function. Of course, 
in a chaotic 
system, the result must be analyzed more deeply since the phase space 
volume remains finite, which leads to a saturation of entropy
even with no dissipation.

We finally point out a very simple inversion formula which
allows one to retrieve the initial state of the system, which seems
a very transparent way to deal with  the topics of quantum state 
reconstruction.

We thank Luiz Davidovich, Ruinet Matos Filho and 
Fabricio Toscano for helpful
discussions. We acknowledge financial support from Faperj, CNPq, Pronex
and the Instituto do Mil\^enio for Quantum Information.

\appendix

\section{Lindblad equation in the chord space}

From the definition (\ref{chordtrans}) of the chord transform,
\be
\tW(\vxi) = \frac{1}{2\pi\hbar} \int d\x ~ 
\exp{ \left( -\frac{i}{\hbar}\vxi\wedge\x \right) } W(\x),
\ee
applied to the derivatives of $W$ and the product of $W$ with $q$ or $p$,
we set the following transformation rules by using integration by parts:
\begin{eqnarray}
W & \rightarrow & \tW \cr
\frac{\der W}{\der \x} & \rightarrow & \frac{i}{\hbar} \J\vxi \tW \cr
\x W & \rightarrow &  - \frac{\hbar}{i} \J \frac{\der \tW}{\der \vxi} \cr
\x\cdot\frac{\der W}{\der \x} & \rightarrow & -2\tW - 
\vxi\cdot\frac{\der \tW}{\der \vxi}.
\label{rules-chords}
\end{eqnarray}
By applying these rules on the following Poisson bracket, 
\be
\left\{ H(\x), W(\x) \right\} = 2 ( \mH_{12} p + \mH_{22} q ) 
\frac{\der W}{\der p}(\x)
- 2 ( \mH_{11} p + \mH_{12} q ) \frac{\der W}{\der q}(\x),
\ee
we get
\be
\left\{ H(\x), W(\x) \right\} \rightarrow 
2 ( \mH_{12} \xi_p + \mH_{22} \xi_q ) \frac{\der \tW}{\der \xi_p}(\x)
- 2 ( \mH_{11} \xi_p + \mH_{12} \xi_q ) \frac{\der \tW}{\der \xi_q}(\x),
\ee
that is
\be
\left\{ H(\x), W(\x) \right\} \rightarrow \left\{ H(\vxi), \tW(\vxi) \right\}.
\ee
In the same way we get
\be
\lambda^2 \frac{\der^2 W}{\der q^2}(\x) + \mu^2 \frac{\der^2 W}{\der p^2}(\x)
-2\lambda\mu \frac{\der^2 W}{\der p \der q}(\x) \rightarrow
- \frac{1}{\hbar^2} \left( \vl\cdot\vxi \right)^2 \tW,
\ee
hence the final equation (\ref{lindchord-genDK}) for $\tW$, 
with the help of the last line of (\ref{rules-chords}).

\section{Zeroes of the Husimi function}

Defining the complex variable $z(\x)=(q+ip)/\sqrt{2\hbar}$, we may express 
the coherent state $|z\rangle$ \cite{MeySar:book} as
\be
|z\rangle = e^{-|z|^2/2} \sum_{n=0}^{\infty} \frac{z^n}{\sqrt{n!}}|n\rangle,
\ee
where $|n\rangle$ are the eigenstates of the harmonic oscillator.
Hence, the coherent state representation of any pure state $|\psi\rangle$
can be expressed as
\be
\langle z|\psi\rangle = e^{-|z|^2/2} F(z^*),
\ee
where
\be
F(z) = \sum_{n=0}^{\infty}\frac{z^n}{\sqrt{n!}}\langle n | \psi\rangle
\ee
is an entire function, known as the Bargmann function \cite{Bar67}.
Since the Husimi function is the square modulus of the coherent state
representation, we obtain
\be
Q(\x) = |\langle z(\x) | \psi \rangle |^2 = e^{-|z|^2}|F(z)|^2.
\label{defhusimi}
\ee

Thus the zeroes of the Husimi function coincide with the zeroes of an
analytic function. Indeed, we may define the state $|\psi\rangle$ by its
Bargmann representation $F(z)$. An important a priori restriction is that
$F(z)$ is at most of order $r=2$.

 To see this, recall that $F(z)$ is of finite order if
\be
|F(z)|<e^{|z|^\mu}
\label{deforder}
\ee
for all sufficiently large $|z|$ and the order of this function is
$r = \inf{\mu}$ for which (\ref{deforder}) holds. But if we use 
the fact that $|\langle z|\psi\rangle|^2\leq 1$ in (\ref{defhusimi}),
we obtain
\be
|F(z)|\leq e^{|z|^2/2}<e^{|z|^2},
\ee
so $r\leq 2$.

We now make use of the following 

Theorem \cite{Mar:book}: If $F(z)$ is an entire function
of finite order with no zeroes on the plane, then its order is necessarily 
an integer and $F(z)=e^{P_{r}(z)}$ where $P_{r}(z)$ is a
polynomial of order $r$.

In the case of the Bargmann function, then $P_{r}(z)$ is at most of second
order and hence $Q(\x)$ given by (\ref{defhusimi}) must be a Gaussian
in $p$ and $q$ if it represents a normalized function.

Thus, only Gaussian Husimi functions have no zeroes in the phase plane.
To a great extent, positions of the isolated zeroes of the Husimi function
also restrict the class of admissible pure states through the factorization
theorems of Weierstrass and Hadamard \cite{Mar:book}. The characterization of the pure
states as chaotic or regular by the pattern of zeroes has been extensively
studied for the case where the phase space is a torus, because the
restriction is then more severe \cite{LebVor90}.



\end{document}